%% file: charm2015_SabatoLeo.tex
\documentclass[12pt]{article}
\usepackage{graphicx}
\usepackage[percent]{overpic}
\usepackage{hyperref}

\newcommand\pubnumber{WSU--HEP--XXYY}
\newcommand\pubdate{\today}

\def\illinois{Department of Physics\\
University of Illinois at Urbana-Champaign, Urbana, IL 61801, USA}
\def\support{\footnote{On behalf of the CDF collaboration.}}

\def\Title#1{\begin{center} {\Large #1 } \end{center}}
\def\Author#1{\begin{center}{ \sc #1} \end{center}}
\def\Address#1{\begin{center}{ \it #1} \end{center}}

\newcommand\pubblock{\rightline{\begin{tabular}{l} \pubnumber\\
         \pubdate  \end{tabular}}}
\newenvironment{Abstract}{\begin{quotation}  }{\end{quotation}}
\newenvironment{Presented}{\begin{quotation} \begin{center} 
             PRESENTED AT\end{center}\bigskip 
      \begin{center}\begin{large}}{\end{large}\end{center} \end{quotation}}

\input econfmacros.tex

\newcommand{\CP}{{\ensuremath{C\!P}}}
\newcommand{\Acp}{\ensuremath{\mathcal{A}_\CP}}

\newcommand{\stat}{\ensuremath{\mathrm{~(stat)}}}
\newcommand{\syst}{\ensuremath{\mathrm{~(syst)}}}
\newcommand{\Dbar}{\ensuremath{\overline{D}{}}}

\newcommand{\tev}{\ensuremath{\mathrm{Te\kern -0.1em V}}}
\newcommand{\gev}{\ensuremath{\mathrm{Ge\kern -0.1em V}}}	
\newcommand{\mev}{\ensuremath{\mathrm{Me\kern -0.1em V}}}	
\newcommand{\kev}{\ensuremath{\mathrm{ke\kern -0.1em V}}}	
\newcommand{\massgev}{\mbox{\gev/$c^2$}}			
\newcommand{\massmev}{\mbox{\mev/$c^2$}}			
\newcommand{\pgev}{\mbox{\gev/$c$}}				
\newcommand{\pmev}{\mbox{\mev/$c$}}				

\begin{document}
\begin{titlepage}
\pubblock

\vfill
\Title{CDF results on \CP\ violation in charm\\ \vspace*{2.0cm}}
\vfill
\Author{Sabato Leo\support}
\Address{\illinois}
\vfill
\begin{Abstract}
I discuss the measurement of \CP-violating asymmetries ($A_{\Gamma}$) between effective lifetimes of $D^0$ or $\Dbar^0$ mesons. Fully reconstructed \mbox{$D^0\to K^+ K^-$} and \mbox{$D^0\to \pi^+\pi^-$} decays collected in $p\bar{p}$ collisions by the Collider Detector at Fermilab experiment and corresponding to a data set of $9.7$~fb$^{-1}$ of integrated luminosity are used. The flavor of the charm meson at production is determined by exploiting the decay $D^{*+} \to D^0 \pi^+$. Contamination from mesons originated in $b$-hadron decays is subtracted from the sample. Signal yields as functions of the observed decay-time distributions are determined using likelihood fits and used to measure the asymmetries. The results, $A_\Gamma (K^+K^-) = \bigl(-1.9 \pm 1.5 \stat \pm 0.4 \syst\bigr)\times10^{-3}$ and $A_\Gamma (\pi^+\pi^-)= \bigl(-0.1 \pm 1.8 \stat \pm 0.3 \syst\bigr)\times10^{-3}$, and their combination, $A_\Gamma = \bigl(-1.2 \pm 1.2)\times10^{-3}$, are consistent with the SM predictions and other experimental determinations. 

\end{Abstract}
\vfill
\begin{Presented}
The 7th International Workshop on Charm Physics (CHARM 2015)\\
Detroit, MI, 18-22 May, 2015
\end{Presented}
\vfill
\end{titlepage}
\def\thefootnote{\fnsymbol{footnote}}
\setcounter{footnote}{0}

\section{Introduction}

In the SM \CP\ violation in charm decays is predicted to be negligibly small, since the dynamics of these decays, at leading order, only involves the first two quark generations~\cite{theory}. Indeed, no \CP-violating effects have been experimentally established yet in charm dynamics~\cite{hfag}.\par
%
Decay-time-dependent rate asymmetries of decays into \CP\ eigenstates, such as $D \to h^+h^-$, where $D$ indicates a $D^0$ or $\overline{D}^0$ meson, and $h$ a $K$ or $\pi$ meson, are sensitive probes for \CP\ violation~\cite{SCS-decays}. Such asymmetries,
\begin{equation}\label{eq:acp}
\Acp(t) = \frac{d\Gamma(D^0\to h^+h^-)/dt - d\Gamma(\Dbar^0\to h^+h^-)/dt}{d\Gamma(D^0\to h^+h^-)/dt+d\Gamma(\Dbar^0\to h^+h^-)/dt},
\end{equation}
\noindent probe non-SM physics contributions in the {\it oscillation} and {\it penguin} transition amplitudes. Either amplitude may be affected by the exchange of non-SM particles, which could enhance the magnitude of the observed \CP\ violation with respect to the SM expectation. The asymmetry $\Acp(t)$ thus receives contributions direct \CP\ violation and from indirect \CP\ violation. Because of the slow oscillation rate of charm mesons~\cite{hfag}, Eq.~(\ref{eq:acp}) is approximated to first order as \cite{AcpCDF},
\begin{equation}\label{eq:acp3}
\Acp(t) \approx \Acp^{\rm{dir}}(h^+h^-) - \frac{t}{\tau}\ A_\Gamma(h^+h^-),
\end{equation}
\noindent where $t$ is the proper decay time 
and $\tau$ is the \CP-averaged $D$ lifetime~\cite{pdg}. The first term arises from direct \CP\ violation and depends 
on the decay mode; the second term is proportional to the asymmetry between the {\it effective} lifetimes $\hat{\tau}$ of anticharm and charm mesons,
\begin{equation}
A_{\Gamma} = \frac{\hat{\tau}(\Dbar^0\to h^+h^-)-\hat{\tau}(D^0\to h^+h^-)}{\hat{\tau}(\Dbar^0\to h^+h^-)+\hat{\tau}(D^0\to h^+h^-)},
\end{equation}
\noindent and is mostly due to indirect \CP\ violation. Effective lifetimes are defined as those resulting from a single-exponential fit of the time evolution of neutral meson decays that may undergo oscillations. In the SM, $A_{\Gamma}$ is universal for all final states with same \CP-parity~\cite{Grossman:2006jg}, such as $K^+K^-$ and $\pi^+\pi^-$; contributions from non-SM processes may introduce channel-specific differences. 
To date, all experiments report $A_{\Gamma}$ values consistent with \CP\ symmetry at the $\mathcal{O}(10^{-3})$ level~\cite{hfag}. The Belle and BaBar collaborations combined the results obtained in  $K^+K^-$ and $\pi^+\pi^-$ \CP--even final states, reporting $A_{\Gamma} = (-0.3 \pm 2.0 \pm 0.8)\times 10^{-3}$ and $A_{\Gamma} = (0.9 \pm 2.6 \pm 0.6)\times 10^{-3}$, respectively~\cite{BfactoriesAgamma}. The most precise results have been reported by the LHCb collaboration separately for the two channels: 
$A_{\Gamma}(\pi^+\pi^-) = (0.33 \pm 1.06 \pm 0.14)\times 10^{-3}$ and $A_{\Gamma}(K^+K^-) = (-0.35 \pm 0.62 \pm 0.12)\times 10^{-3}$~\cite{LHCbAgamma}.  All the above results, along with more recents results from the LHCb experiment~\cite{LHCb_muontagged}, are consistent with \CP\ symmetry with $\mathcal{O}(10^{-3})$ uncertainties. 
\par Singly--Cabibbo--suppressed decays into \CP--eigenstates, such as $D^0\to\pi^+\pi^-$ and $D^0\to K^+K^-$ are convenient channels for pursuing a measurement of lifetime asymmetry. Their final states can be fully reconstructed, providing a precise determination of the decay time, and the decays have significant signal yields and moderate backgrounds, allowing for reduced systematic uncertainties. With the full data set of $9.7$\,fb$^{-1}$ of data, CDF aims to a sensitivity of $\mathcal{O}(10^{-3})$, comparable with other experiments' sensitivities. While the decay-time distribution is biased by the online selection on long-lived decays and transverse decay lengths, the effect of the bias cancels to a high level of accuracy in the asymmetry between distributions associated with the same final state and any residual effects can be checked against in control samples with similar kinematic properties.
In order to extract $A_{\Gamma}$, we determine separately the yields of primary $D^0\to h^+h^-$ and $\Dbar^0\to h^+h^-$ decays as functions of reconstructed $D$ decay time. The analysis uses only candidates populating a narrow range centered around the known $D^0$ meson mass. The flavor at production is identified by the charge of the low-momentum pion (soft pion, $\pi_s$) in the strong-interaction decay $D^{\star +}\to D^0\pi^+$. Each sample is divided into subsamples according to production flavor and decay time. In each subsample, a fit to the $D\pi_s^{\pm}$ mass distribution is used to determine the relative proportions of signal and background. 
The results of these fits are used to construct a background-subtracted distribution of the $D$ impact parameter, the minimum distance from the beam of the $D$ trajectory. This distribution is fit to identify  $D^{*\pm}$ mesons from $b$-hadron decays ({\it secondary decays}), whose observed decay-time distribution is biased by the additional decay length of the $b$-hadron, and to determine the yields of charm ($N_{D^{0}}$) and anticharm ($N_{\overline{D}^{0}}$) mesons directly produced in the $p\bar{p}$ collision ({\it primary decays}). The yields are combined into the asymmetry $A=(N_{D^0}-N_{\Dbar^0})/(N_{D^0}+N_{\Dbar^0})$, which is fit according to Eq.~(\ref{eq:acp3}). The slope yields $A_{\Gamma}$. The intercept determines the asymmetry at $t=0$, $A(0)$, which receives contributions from direct \CP\ violation and possible instrumental asymmetries. 
The size of a possible decay-time dependence of the detector asymmetry is constrained using large control samples of $13\times10^{6}$ $D^{*\pm}\to D(\to K^\mp\pi^\pm)\pi_s^\pm$ signal decays where \CP\ violation is negligible, if any. Sample selection, studies of background composition, and fit model follow from previous measurements~\cite{AcpCDF}.\par

\section{The CDF II detector}
The CDF II detector is a multipurpose magnetic spectrometer surrounded by calorimeters and muon detectors. The detector components relevant for this analysis are outlined as follows; a detailed description is in Ref.~\cite{CDF}. A silicon microstrip vertex detector and a cylindrical drift chamber immersed in a $1.4$~T axial magnetic field allow reconstruction of charged-particle trajectories (tracks) in the pseudorapidity range $|\eta| < 1.0$. The vertex detector contains seven concentric layers of single- and double-sided silicon sensors at radii between 1.5 and 22~cm, each providing a position measurement with up to 15 (70) $\mu$m resolution in the $\phi$ ($z$) direction~\cite{silicon}. The drift chamber has 96 measurement layers, between 40 and 137~cm in radius, organized into alternating axial and $\pm 2^{\circ}$ stereo superlayers~\cite{COT}. The component of a charged particle's momentum transverse to the beam ($p_T$) is determined with a resolution of $\sigma_{p_T}/p_T^2 \approx 0.07\%\ (\pgev)^{-1}$, corresponding to a typical mass resolution of $8~\massmev$ for a two-body charm-meson decay.
\par The data are collected by a three-level trigger. At level 1, hardware custom processors reconstruct tracks in the transverse plane of the drift chamber. Two oppositely-charged particles are required, with reconstructed transverse momenta $p_{T} > 2~\pgev$, scalar sum $\sum p_{T} > 5.5~\pgev$ and azimuthal opening angle $\Delta \phi < 90^{\circ}$~\cite{XFT}. At level 2, tracks are combined with silicon hits and their impact parameter (transverse distance of closest approach to the beam line) is determined with 45~$\mu$m resolution (including the beam spread) and typically required to be between 0.12 and 1.0 mm~\cite{SVT}. A more stringent opening-angle requirement of $2^{\circ}<\Delta\phi<90^{\circ}$ is also applied. Each track pair is then used to form a $D$ candidate, whose flight distance in the transverse plane projected onto the transverse momentum ($L_{xy}$) is required to exceed $200~\mu$m. At level 3, the selection is reapplied on events fully reconstructed by an array of processors.\par

\section{Selection and reconstruction}

\begin{figure}[t]
\centering
\includegraphics[width=3in]{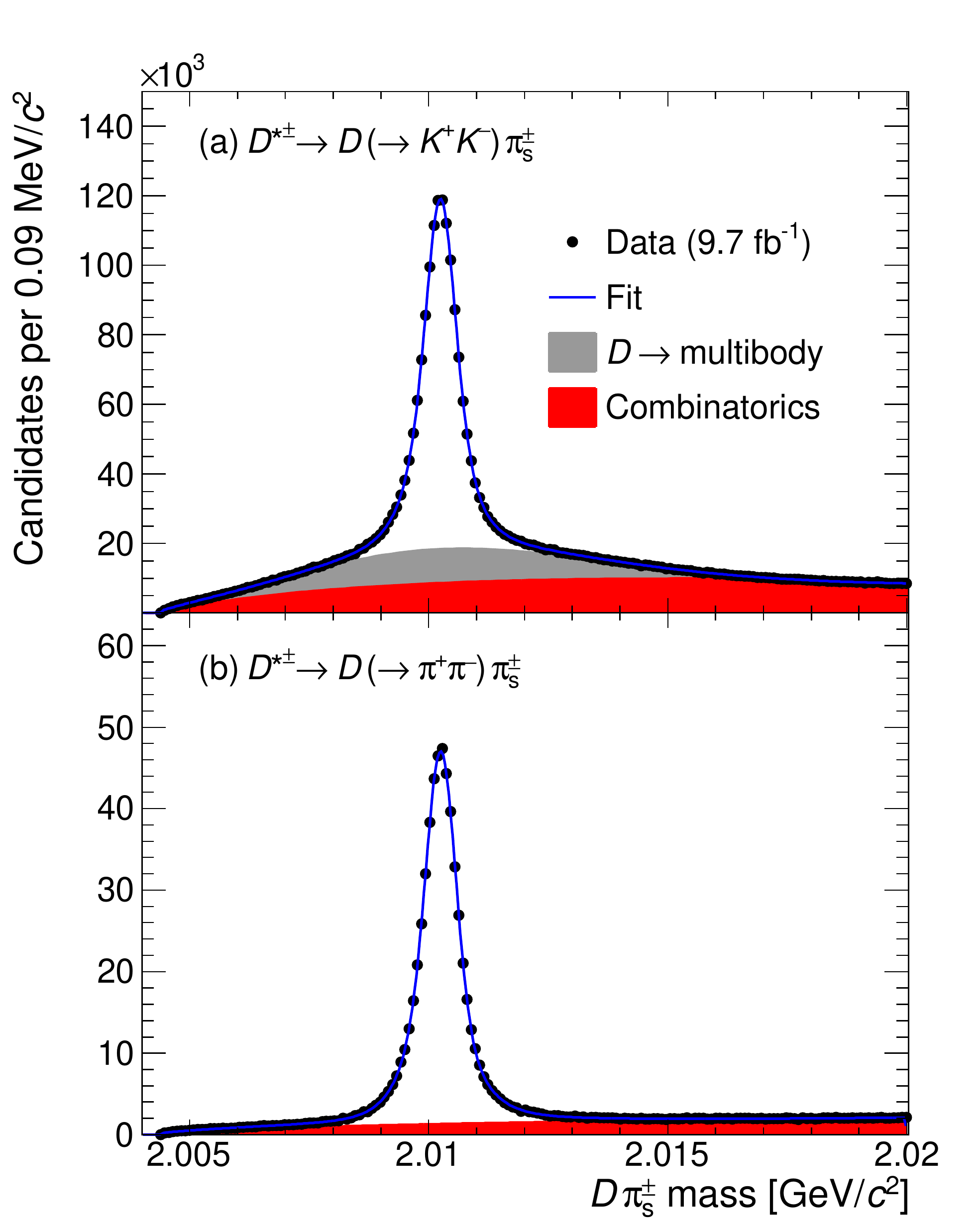}
\caption{Distributions of $D\pi^{\pm}$ mass with fit results overlaid for (a) the $D \to K^+K^-$ and (b) $D \to \pi^+\pi^-$ sample.}\label{fig:mass}
\end{figure}

Online data selection is based on pairs of charged particles displaced from the $p\bar{p}$ collision point. Offline, a $D$ candidate is reconstructed using two oppositely charged tracks fit to a common decay vertex. A charged particle with $p_T>400~\pmev$ is associated with each $D$ candidate to form $D^{*\pm}$ candidates. 
Constraining the $D^{*\pm}$ decay vertex to lie on the beam-line results in a 25\% improvement in $D^{*\pm}$ mass resolution w.r.t Ref.~\cite{AcpCDF}.  Ref.~\cite{AcpCDF} details the offline selection. The $h^+h^-$ mass of selected candidates is required to be within about 24 \massmev\ of the known $D^0$ mass, $m_{D^0}$~\cite{hfag}, to separate $D \to K^+K^-$ and $D \to \pi^+\pi^-$ samples.
Final selected samples contain $6.1\times 10^5$ $D^0 \to K^+K^-$,  $6.3\times 10^5$ $\overline{D}^0 \to K^+K^-$, $2.9\times 10^5$ $D^0 \to \pi^+\pi^-$, and $3.0\times 10^5$ $\overline{D}^0 \to \pi^+\pi^-$ signal events. The main backgrounds are real $D^0$ decays associated with random pions or random combinations of three tracks (combinatorics) for the $\pi^+\pi^-$ sample, while the $K^+K^-$ sample is also polluted by misreconstructed multibody charm meson decays (i.e. $D^0 \to h^-\pi^+ \pi^0$ and $D^0 \to h^- \ell^+ \nu_{\ell}$, where $\ell$ is a muon or an electron), see Figure~\ref{fig:mass}. 

\begin{figure}[t]
\centering
\includegraphics[width=3in]{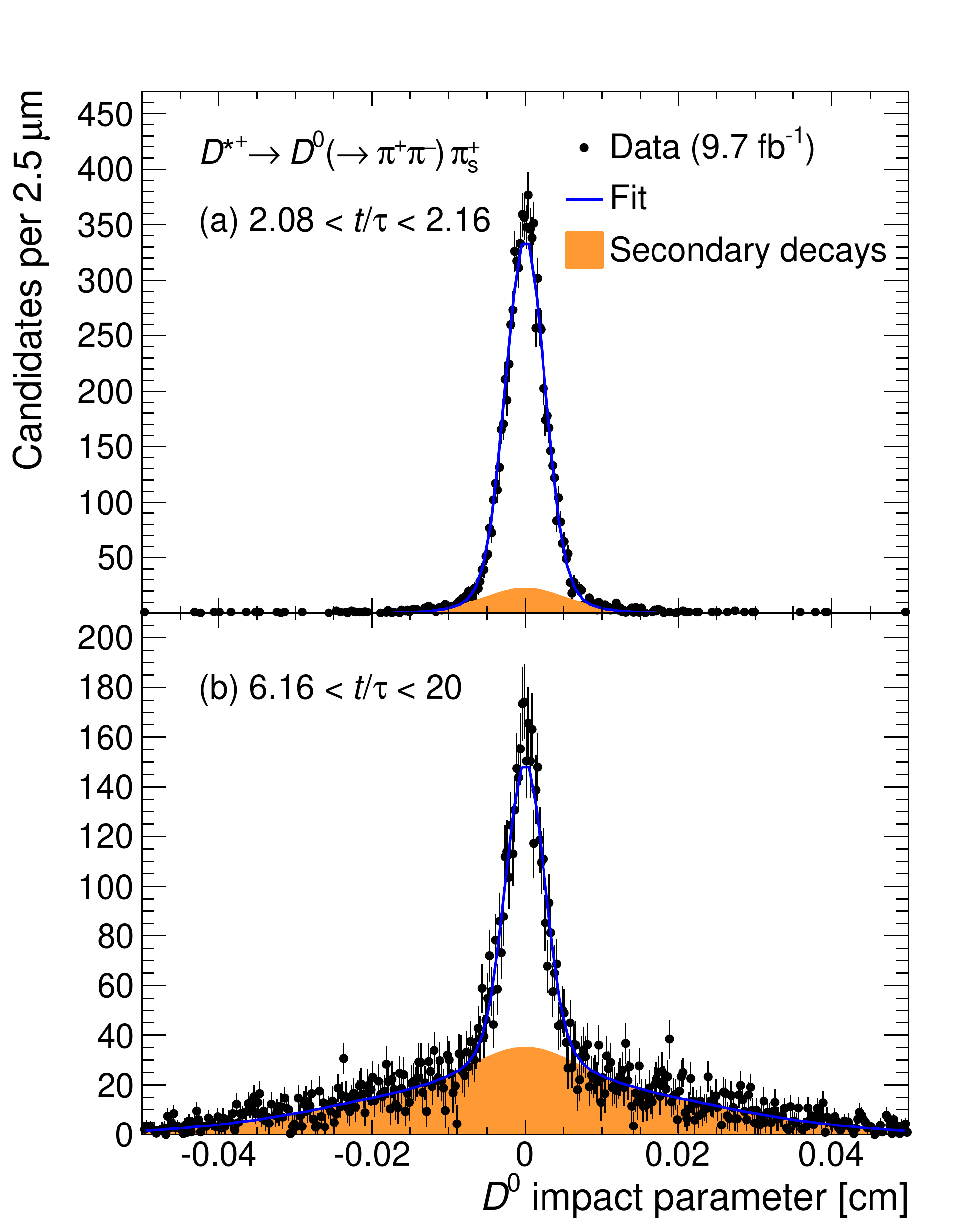}
\caption{Distributions of $D^0$ impact parameter with fit results overlaid for background-subtracted $D^{*+} \to D^0 (\to \pi^+\pi^-)\pi_s^+$ decays restricted to (a) the decay-time bin $2.08 < t/\tau < 2.16$ and (b) the decay-time bin $6.16 < t/\tau < 20$.}\label{fig:ip_fits}
\end{figure}

\section{Determination of the asymmetry}

\begin{figure}[htb]
\centering
\includegraphics[height=3in]{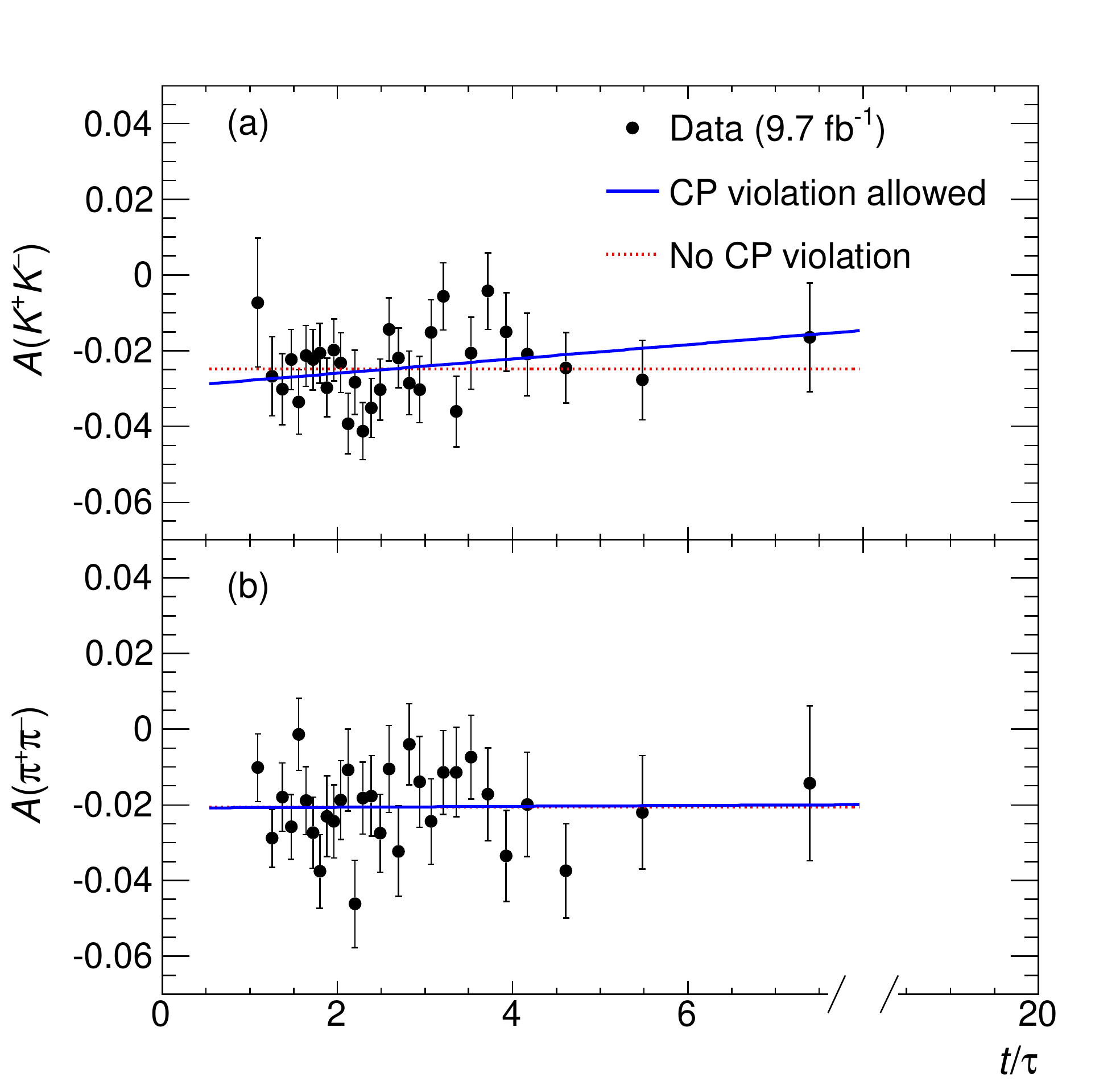}
\caption{Effective lifetime asymmetries as functions of decay time for the (a) $D \to K^+K^-$ and (b) $D \to\pi^+\pi^-$ samples. Results of fits not allowing for (red dotted line) and allowing for (blue solid line) \CP\ violation are overlaid.}\label{fig:Agamma}
\end{figure}

The flavor-conserving strong-interaction processes $D^{*+} \to D^0 \pi^+$ and $D^{*-} \to \Dbar^0\pi^-$ allow identification of the initial flavor through the charge of the low-momentum $\pi$ meson (soft pion, $\pi_s$).  $D^0$ or $\Dbar^0$ subsamples are thus divided in equally populated 30 bins of decay time between $0.15\tau$ and $20\tau$. 
In each bin, the average decay-time $\vev{t}$ is determined using $13\times10^{6}$ $D^{*\pm}\to D(\to K^\mp\pi^\pm)\pi_s^\pm$ signal decays. 
Signal and background yields in the signal region are determined in each decay-time bin, and for each flavor, through $\chi^2$ fits of the $D\pi_{s}^{\pm}$ mass distribution. 
The model for the the signal shapes is determined from simulation~\cite{AcpCDF}, with parameters tuned in the sample of $D\to K^\mp\pi^\pm$ decays, independently for each $D$ flavor and decay-time bin. The resulting signal-to-background proportions are used to construct signal-only distributions of the $D$ meson impact parameter (IP). In each bin and for each flavor, background-subtracted IP distributions are formed by subtracting IP distributions of background candidates, sampled in the $2.015 < M(D\pi^{\pm})<2.020~\massgev$ region for the $\pi^+\pi^-$ sample, from IP distributions of signal candidates which have $M(D\pi_s^\pm)$ within $2.4~\massmev$ of the known $D^{*+}$ mass~\cite{hfag}. Contamination from multibody decays in the $K^+K^-$ sample is accounted for by using candidates in the sideband $m_{D^0}-64~\massmev<M(K^+K^-)<m_{D^0}-40~\massmev$ and with $M(D\pi_s^\pm)$ within $2.4~\massmev$ of the known $D^{*\pm}$ mass. 
A $\chi^2$ fit of these signal-only IP distributions identifies $D^{*\pm}$ mesons from $b$-hadron decays ({\it secondary})
and determines the yields of charm ($N_{D^{0}}$) and anticharm ($N_{\overline{D}^{0}}$) mesons directly produced in the $p\bar{p}$ collision ({\it primary}). 
Double-Gaussian models are used for both the primary and secondary components. The parameters of the primary component are derived from a fit of candidates in the first decay-time bin ($t/\tau<1.18$), where any bias from the $\mathcal{O}(\%)$ secondary contamination is negligible, and fixed in all fits. 
The parameters of the secondary component are determined by the fit independently for each decay-time bin, see Figure~\ref{fig:ip_fits}.
The yields are then combined into the asymmetry $A=(N_{D^0}-N_{\Dbar^0})/(N_{D^0}+N_{\Dbar^0})$, which is fit with the linear function in Eq.\ (\ref{eq:acp3}). 
The slope of the function, which yields $A_{\Gamma}$, is extracted using a $\chi^2$ fit.
The fit is shown in Fig.~\ref{fig:Agamma} and yields $A_{\Gamma}(K^+K^-) = \bigl(-1.9 \pm 1.5\stat\bigr)\times 10^{-3}$ and $A_{\Gamma}(\pi^+\pi^-) = \bigl(-0.1 \pm 1.8\stat\bigr)\times 10^{-3}$. In both samples, we observe few percent values for $A(0)$, due to the known detector-induced asymmetry in the soft-pion reconstruction efficiency~\cite{AcpCDF}. The independence of instrumental asymmetries from decay time is demonstrated by the analysis of $D \to K^\mp\pi^\pm$ decays, where no indirect \CP\ violation occurs and instrumental asymmetries are larger; an asymmetry compatible with zero is found, $(-0.5 \pm 0.3)\times 10^{-3}$. \par
Table~\ref{tab:syst} summarizes the most significant systematic uncertainties.

\begin{table}[h!]
\centering
\begin{tabular}{lcc}
\hline\hline
Source  					& $\Delta A_\Gamma(\pi^+\pi^-)$ & $\Delta A_\Gamma(K^+K^-)$\\
\hline
Background subtraction  	& 0.021\% & 0.038\%\\
Impact parameter shapes  & 0.026\% & 0.010\%\\
Decay-time scale         & 0.001\% & 0.003\%\\
\hline
Total 					& 0.033\% & 0.039\%\\
\hline\hline
\end{tabular}
\caption{Summary of most significant systematic uncertainties. The total uncertainty is the sum in quadrature of all the contributions.}\label{tab:syst}
\end{table}

For the $\pi^+\pi^-$ analysis, the dominant systematic uncertainty of 0.028\% arises from the choice of the impact-parameter shape of the secondary component whereas for the $K^+K^-$ sample this effect only contributes 0.013\%. The choice of the background sideband has a dominant effect in the $K^+K^-$ analysis (0.038\%) and a minor impact (0.010\%) on the $\pi^+\pi^-$ result. 

\par

\section{Conclusions}
The CDF experiment pioneered the investigation of \CP\ violation in the charm sector with a successful program that spans over more than 10 years and established that world-leading charm physics is possible at hadron colliders.\par
In 2005 a pioneering measurement of direct \CP\ violation in $D^0 \to h^+h^-$ decays~\cite{cdf_prl94} yielded no evidence of \CP\ violation. In 2012 we measure \CP\ violating asymmetries in  $D^0\to\pi^+\pi^-$ and $D^0\to K^+K^-$ decays with about 6 fb$^{-1}$ of CDF run II data~\cite{AcpCDF}. The analysis did not use any Montecarlo input nor assumption on the \CP\ conservation in Cabibbo-favored decays and yielded results consistent with \CP\ conservation and in agreement with theoretical predictions~\cite{AcpCDF}. In 2012 CDF measures the difference in \CP\ violating asymmetries in  $D^0\to\pi^+\pi^-$ and $D^0\to K^+K^-$ decays.\par
The measurement of the difference in effective lifetime between anticharm and charm mesons reconstructed in $D^0 \to K^+K^-$ and $D^0 \to \pi^+\pi^-$ decays using the full CDF data set is then just the latest of such a ``charmingly" successful effort. 
The final results,
\beqa
A_{\Gamma}(\pi^+\pi^-) = (-0.1 \pm 1.8 \stat  \pm 0.3 \syst)\times10^{-3},\CR
A_{\Gamma}(K^+K^-) = (-1.9 \pm 1.5 \stat  \pm 0.4 \syst)\times10^{-3},
\eeqan
are consistent with \CP\ symmetry and combined to yield $A_\Gamma = \bigl(-1.2 \pm 1.2)\times10^{-3}$~\cite{Aaltonen:2014efa}. The results are also consistent with the current best results \cite{BfactoriesAgamma,LHCbAgamma} and contribute to improve the global constraints on indirect \CP\ violation in charm meson dynamics.



\end{document}

%% file: econfmacros.tex
\def\beq{\begin{equation}}
\def\eeq#1{\label{#1}\end{equation}}
\def\eeqn{\end{equation}}

\def\beqa{\begin{eqnarray}}
\def\eeqa#1{\label{#1}\end{eqnarray}}
\def\eeqan{\end{eqnarray}}
\def\CR{\nonumber \\ }

\let\bar=\overbar

\def\vev#1{\langle #1 \rangle}

\def\Dslash{\not{\hbox{\kern-4pt $D$}}}
\def\dslash{\not{\hbox{\kern-2pt $\del$}}}

\def\msb{{\bar{\ssstyle M \kern -1pt S}}}

